\documentclass{moriond}

\def\be{\begin{equation}}
\def\ee{\end{equation}}
\def\bea{\begin{eqnarray}}
\def\eea{\end{eqnarray}}

\newcommand{\Photo}{}

\usepackage{siunitx}
\newcommand{\asym}[3]{\ensuremath{\num{#1}^{\num[print-implicit-plus = true]{#2}}_{\num{-#3}}}}
\usepackage{lineno}
\nolinenumbers

\begin{document}
\vspace*{4cm}
\title{Searches for Galactic Neutrinos with the IceCube Neutrino observatory}

\author{A. Sandrock on behalf of the IceCube collaboration\footnote{%
\protect\url{http://icecube.wisc.edu}}}

\address{Department of Mathematics and Natural Sciences, University of Wuppertal,\\
Gau\ss{}stra\ss{}e 20, 42119 Wuppertal, Germany}

\maketitle\abstracts{
The sources of galactic charged cosmic rays are so far unknown, because
their arrival directions are randomized in the galactic magnetic field.
Objects accelerating hadrons are expected to produce high-energy
neutrinos. In addition, a diffuse galactic neutrino flux is predicted
from interactions of galactic cosmic rays with matter during propagation
through the galaxy.
The IceCube neutrino observatory at the geographic South Pole
instruments a cubic kilometer of ice with optical modules to detect the
Cherenkov light of particles produced in neutrino interactions.
Operating for more than a decade in its complete detector configuration,
IceCube is in a unique position to search for neutrino sources.
This contribution discusses the searches for a diffuse flux of neutrinos
as wells as for neutrinos from candidate point sources and extended
sources in the galactic plane.
}

\section{Introduction}
Astrophysical sources can emit various stable messenger particles. While
charged cosmic rays are deflected by magnetic fields, gamma rays and neutrinos
as  neutral particles are undeflected and point back to their sources. Gamma
rays can be produced by either leptonic processes such as synchrotron radiation,
Compton scattering or bremsstrahlung, or by hadronic processes resulting in the
production of mesons. Neutrinos on the other hand are only produced in sources
which are hadron accelerators.

Our own galaxy is a prominent source in photons. These photons can be produced
either in discrete sources or diffusely by galactic charged cosmic rays
via hadronuclear interactions producing mesons. Searching for neutrinos from
our galaxy has the potential to identify hadronic accelerators and thus the
sources of galactic cosmic rays.

The IceCube neutrino observatory\cite{icecube_detector} observes mostly two different event
topologies, track-like events from muons produced in charged current
interactions of muon neutrinos or from atmospheric muons produced in
cosmic-ray induced extensive air showers, and cascade-like events from
charged-current interactions of electron and tau neutrinos or from neutral-%
current interactions of all neutrino flavors.\footnote{At very high energies,
an additional double cascade event signature from charged current tau neutrino
interactions is observed \cite{icecube_tau}.} The direction of track-like events
can be reconstructed with good accuracy of $\sim \SI{1}{\degree}$ at \si{TeV}
energies, while the direction of cascade-like events can only be reconstructed
within $\sim \SI{15}{\degree}$. On the other hand, the energy reconstruction of
cascade-like events is essentially calorimetric and significantly more precise
($\sim 10\%$ on the energy) than the energy reconstruction of through-going
track-like events based on the energy loss profile ($\sim 10\%$ on
$\log(E/\si{GeV})$). Through-going track-like events have the advantage of a
larger effective volume, since they can interact outside the detector and enter
it, while cascade-like events have to interact within the detector itself.

The background for neutrino searches in underground neutrinos telescopes such
as IceCube consists of atmospheric muons and neutrinos. Being located at the
South Pole, the earth can be used as a filter for searches for neutrinos coming
from the northern sky, while at the southern sky neutrinos have to be
distinguished from the atmospheric muon background by searching for events
starting in the detector. The atmospheric neutrino background is mostly due
to muon neutrinos, whereas the electron neutrino flux from the atmosphere is
significantly smaller.

Neutrinos from astrophysical sources and from atmospheric background differ
in their distributions in position on the sky and in their energy spectrum.
The likelihood for a single source can be written as
\begin{equation}
  L = \prod_i^N L_i (\mathbf{\theta}, \mathbf{D}_i) = \prod_i^N \left(
  \frac{n_s}{N} S(\mathbf{\theta}, \mathbf{D}_i) + \frac{N - n_s}{N}
  B(\mathbf{D_i}) \right),
\end{equation}
where $N$ is the total number of neutrinos, $n_s$ the number of source
neutrinos, $\mathbf{\theta}$ the parameters of the source, and $\mathbf{D}_i$
the position on the sky of the $i$-th neutrino event. When searching for
a flux of similar sources, a stacking search is performed, which allows to
search for the total flux of a population of searches too weak to be detected
separately. In this case the likelihood is modified to
\begin{equation}
  L = \prod_i^N \left[\frac{n_s}{N} \left(\sum_j^M w_j S(\mathbf{\theta}_j,
  \mathbf{D}_i) + \frac{N - n_s}{N} B(\mathbf{D}_i)\right) \right],
\end{equation}
where $j = 1, \dots, M$ indexes the sources and $w_j$ denotes a weight for each
source, based on e.\,g. an external criterion such as the gamma-ray flux.

\section{Search for Neutrinos from LHAASO UHE $\gamma$-Ray Sources}
\begin{figure}
  \includegraphics[width=\textwidth]{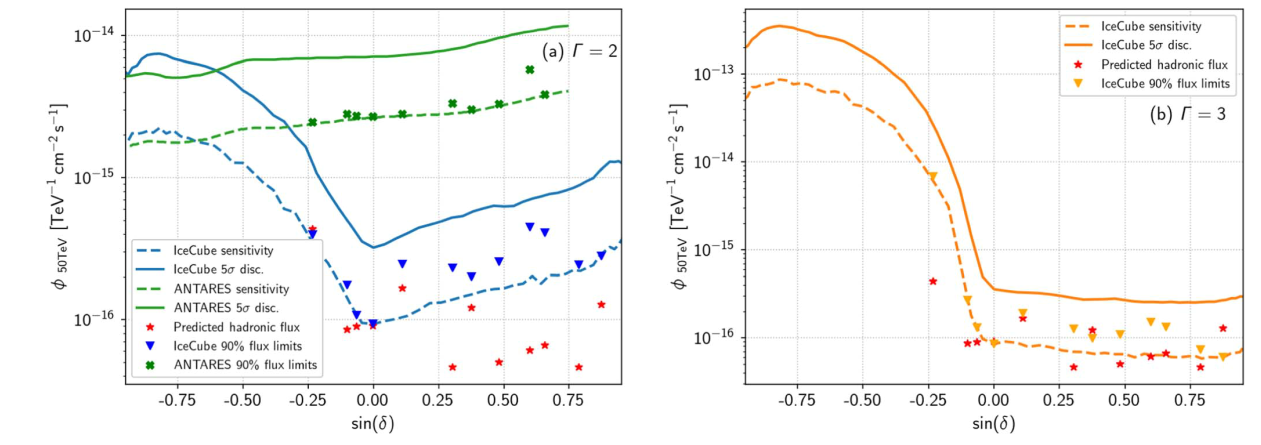}
  \caption{Sensitivity and discovery potential of the search for neutrino
    emission from LHAASO detected UHE $\gamma$-ray sources for spectral indices
    of $\Gamma = 2$ (left) and $\Gamma = 3$ (right). The flux limits determined
    at 90\% C.L. are shown together with the predicted hadronic flux. For
    comparison, the sensitivity and discovery potential for an ANTARES search
    are shown in the left panel. Figure taken from \protect\cite{icecube_lhaaso}.}
  \label{fig:lhaaso_sensitivity}
\end{figure}

The Large High Altitude Air Shower Observatory (LHAASO) has observed 12 sources
with $\gamma$-ray emission above \SI{100}{TeV} \cite{lhaaso_21}. Except the Crab
nebula, none of these has a firm association to known astrophysical objects,
although associations to pulsar wind nebul\ae\ and supernova remnants have been
suggested. At these energies, leptonic emission processes are unlikely due to
the Klein-Nishina cutoff, making these sources good candidates for PeVatron
searches. Using 11 years of IceCube track-like neutrino events from June 2008 to
August 2020, both a catalog search for the single sources as well as stacking
searches for an emission from the whole population of sources were carried
out.\cite{icecube_lhaaso} The sensitivity and discovery potential of this
analysis are shown in Figure~\ref{fig:lhaaso_sensitivity}, together with the
corresponding values for an ANTARES search.\cite{Illuminati:2021Mm}

In the catalog search, no significant emission was observed; the two sources with the
lowest $p$-value are LHAASO~J1908+0621 and LHAASO~J2018+3651. The first one is
probably the UHE counterpart of MGRO~J1908+06; it was the galactic source with
the lowest $p$-value in the latest all-sky IceCube point source analysis, using
10 years of data\cite{icecube_10yr}. The second one is a probable counterpart of
MGRO~J2019+37; in addition, the H~\textsc{ii} region Sh~2-104 to the west of
this source hosts several young massive star clusters and has been observed as
a source of diffuse X-ray emission.

In the stacking search, six analyses were carried out for three selections of
sources with two weighting schemes. The three groups of sources were all LHAASO
observed sources, only the sources possibly associated with supernova remnants,
or only the sources possibly associated with pulsar wind nebul\ae. The sources
were either weighted according to the observed $\gamma$-ray flux at \SI{100}%
{TeV} or giving each source in the sample an equal weight. None of the performed
analyses resulted in a significant detection, therefore 90\% C.L. upper limits
on the flux were determined.

The 90\% C.L. upper limits from the catalog and stacking searches were compared
to the neutrino flux predicted from the assumption that all $\gamma$-rays are
produced by hadronic processes\cite{ahler_murase2014}. This allowed us to constrain
the fraction of hadronically produced gamma-rays for the supernova remnant
sample for both weighting schemes and for the pulsar wind nebula sample for the
flux-weighted sample. For the catalog search, the fraction of hadronically
produced photons could be constrained to $< 47\%$ for the supernova remnant
G106.3+02.7 and $<59\%$ for the Crab nebula, assuming an unbroken powerlaw flux.
Assuming a log-parabola spectral shape, the constraint on the Crab nebula
weakens to $<84\%$ and we can no longer constrain the fraction for G106.3+02.7.

\section{Observation of high-energy neutrinos from the Galactic plane}
Searching for a diffuse neutrino flux from the Galactic plane is difficult
using track-like events due to the large background of air-shower induced
muon events. Therefore, a search using cascade-like events was carried out
\cite{icecube_galactic_plane}, since the background is significantly smaller.
The worse angular resolution of cascade events has been improved significantly
by recently developed deep-learning based reconstruction methods
\cite{icecube_jinst}. The search for a diffuse neutrino flux was implemented
as a template search based on pre-existing models of the galactic neutrino
emission, thus leaving one degree of freedom, namely the normalization of the
template flux, leaving its spatial and spectral characteristics constant.

\begin{figure}
  \begin{center}
  \includegraphics[width=0.7\textwidth]{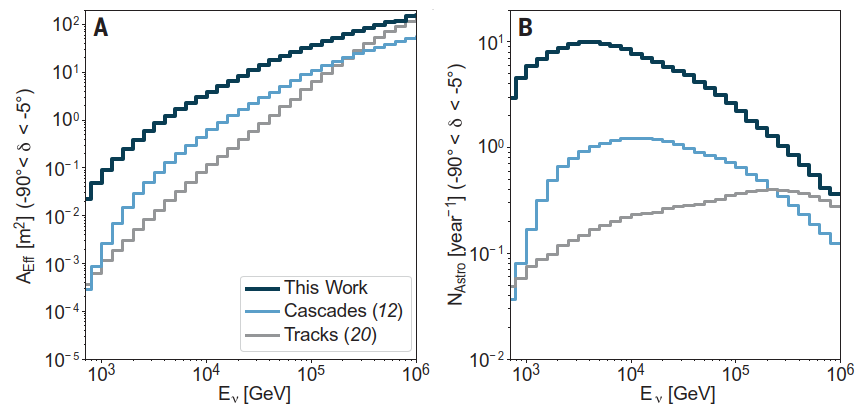}
  \end{center}
  \caption{Effective area (left) and astrophysical neutrino rates (left) of
  the new cascade event selection compared to the previous cascade event
  selection and southern sky track event selection. Figure taken from
  \protect\cite{icecube_galactic_plane}.}
  \label{fig:dnn_cascades}
\end{figure}

While previous event selections were based on high-level observables, for this
search a new event selection based on convolutional neural networks was
developed. This selection uses %observables on the level of the optical modules,
convolutional neural networks,
and its very fast inference speed allows to carry out more complex filtering at
early analysis stages. It keeps significantly more neutrinos at lower energies,
leading to an increase in the effective area equivalent to about 75 years of the
previous cascade event selection (cf. Figure~\ref{fig:dnn_cascades}). This
technology is also applicable to other IceCube analyses, e.\,g. in searches for
exotic particles or special event signatures in cosmic ray analyses.

The templates for the diffuse galactic neutrino flux are based on
\textit{Fermi}-LAT observations. The $\pi^0$ model \cite{ackermann_2012} was
developed based on GALPROP \cite{galprop} simulations to explain the observed
$\gamma$-rays and extrapolates the hadronic component to the energy range of
IceCube. The KRA$\gamma$ model makes different assumptions about the diffusion
coefficient and uses different software for the calculation \cite{gaggero_2015},
resulting in a template with neutrino emission more concentrated on the galactic
center. The KRA$\gamma$ model is tested with two different cutoffs of the proton
spectrum at \SI{5}{PeV} and \SI{50}{PeV}, respectively.

Besides the model-dependent diffuse galactic neutrino search, a catalog stacking
analysis was carried out on a selection of sources from the TeVCat, for the 12
sources with the highest predicted neutrino flux from the following categories:
pulsar wind nebul\ae, supernova remnants, and unidentified sources.
In addition, a template analysis was carried out for neutrino emission from the
Fermi bubbles, and a catalog search for 109 sources from the 4FGL catalog with
the highest $\gamma$-ray flux weighted by sensitivity at the source declination,
most of which are extragalactic.

The galactic plane in optical and $\gamma$-ray photons \cite{gal_optical,%
gal_fermi}, together with the model of the neutrino prediction from the
$\pi^0$ model, the analysis expectation obtained by convolving the template
with detector sensitivity and angular resolution, as well as the significance
map of the experimental results are shown in Figure~\ref{fig:galactic_plane_%
photon_neutrino}.
\begin{figure}
  \includegraphics[width=\textwidth]{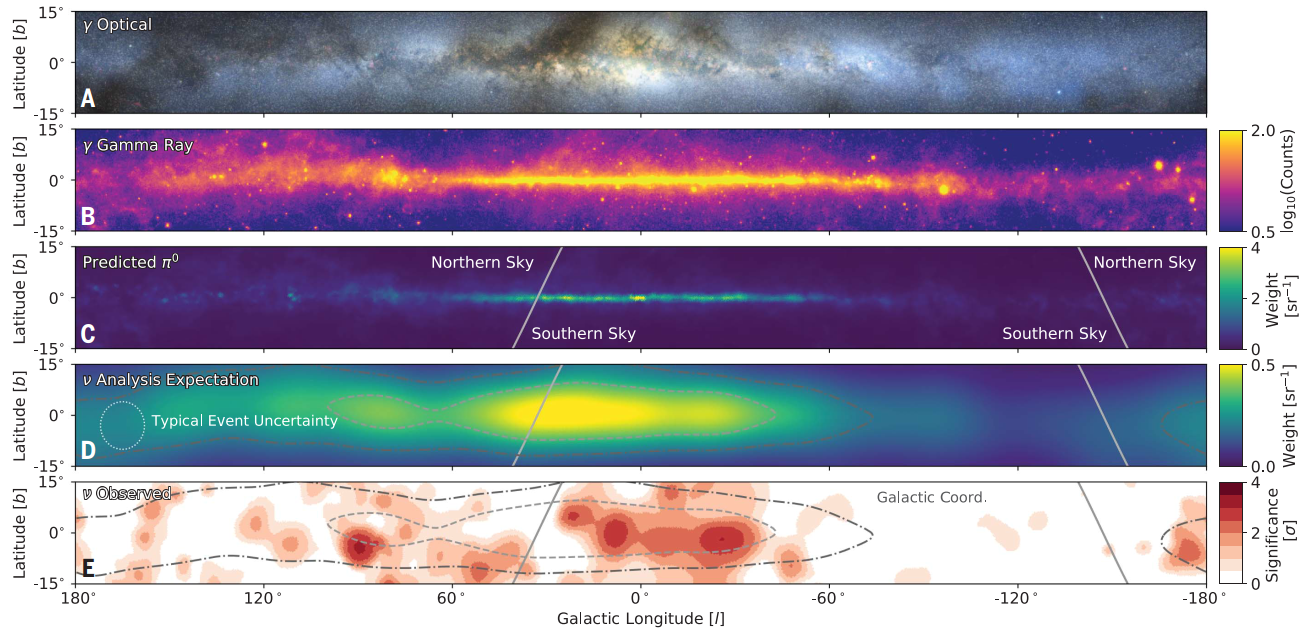}
  \caption{Galactic plane (A) in optical emission, partially obscured by gas and dust;
  (B) in GeV $\gamma$-ray photons as observed by \textit{Fermi}-LAT in 12 years of
  observation, (C) in neutrinos as predicted by the $\pi^0$ model, (D) analysis
  expectation based on the $\pi^0$ model and the detector sensitivity and angular
  reconstruction uncertainty, (E) observed neutrino emission from the Galactic plane.
  Figure taken from \protect\cite{icecube_galactic_plane}.}
  \label{fig:galactic_plane_photon_neutrino}
\end{figure}

\begin{table}
\caption{Results of the galactic plane analysis using cascade-like neutrino
events. The result for the $\pi^0$ model is given as $E^2\ dN/dE$ at \SI{100}%
{TeV} in units of \SI{e-12}{TeV.cm^{-2}.s^{-1}}, for the KRA$\gamma$ models
as multiples of the model flux. Significances denoted with asterisk are
consistent with the template search result and therefore do not denote
independent statistical evidence. Table taken from
\protect\cite{icecube_galactic_plane}.}
\begin{tabular}{lccc}
\hline
 & Flux sensitivity $\Phi$ & $P$ value & Best-fitting flux $\Phi$ \\
\hline
 & & \textit{Diffuse Galactic plane analysis} \\
$\pi^0$ & \num{5.98} & \num{1.26e-6} ($\num{4.71} \sigma$)
  & \asym{21.8}{5.3}{4.9} \\
KRA$_\gamma^5$ & $\num{0.16} \times$ MF & \num{6.13e-6} ($\num{4.37} \sigma$)
  & \asym{0.55}{0.18}{0.15} $\times$ MF \\
KRA$_\gamma^{50}$ & $\num{0.11} \times$ MF & \num{3.72e-5} ($\num{3.96} \sigma$)
  & \asym{0.37}{0.13}{0.11} $\times$ MF \\
 & & \textit{Catalog stacking analysis} \\
SNR & & \num{5.90e-4} $(\num{3.24} \sigma)^*$ \\
PWN & & \num{5.93e-4} $(\num{3.24} \sigma)^*$  \\
UNID & & \num{3.39e-4} $(\num{3.40} \sigma)^*$  \\
 & & \textit{Other analyses} \\
Fermi bubbles & & \num{0.06} $(\num{1.52} \sigma)$ \\
Source list & & \num{0.22} $(\num{0.77} \sigma)$ \\
Hotspot (north) & & \num{0.28} $(\num{0.58} \sigma)$ \\
Hotspot (south) & & \num{0.46} $(\num{0.10} \sigma)$ \\
\hline
\end{tabular}
\label{tab:galactic}
\end{table}
The results of these analyses are given in Table~\ref{tab:galactic}.
The best-fit fluxes for the KRA$\gamma$ models are lower than predicted from
the $\gamma$-ray emission. This could possibly be an indication that the
cutoff of the proton spectrum is inconsistent with the values assumed in
\cite{gaggero_2015}. The best-fit neutrino flux for the $\pi^0$ model, on the
other hand, is about a factor 5 larger than the extrapolation of the flux based
on the observed $\gamma$-ray emission. This might be an indication of spectral
differences or propagation effects in the Galactic center. It is also possible
that a contribution of unresolved point sources is responsible for this
difference.

\section{Search for Extended Sources of Neutrino Emission in the Galactic
Plane with IceCube}
IceCube performed a model-independent search for extended sources of
galactic neutrino \cite{icecube_extended_sources}s, using 9 years of track-%
like neutrino events. Existing surveys of the galactic plane in high-energy
$\gamma$-rays by e.\,g. H.E.S.S.\cite{hess_galactic_plane_survey} and
HAWC\cite{hawc_hess_gp_survey} indicate extended sources with spatial extent up
to $\sim \SI{2}{\degree}$.

\begin{figure}
  \includegraphics[width=\textwidth]{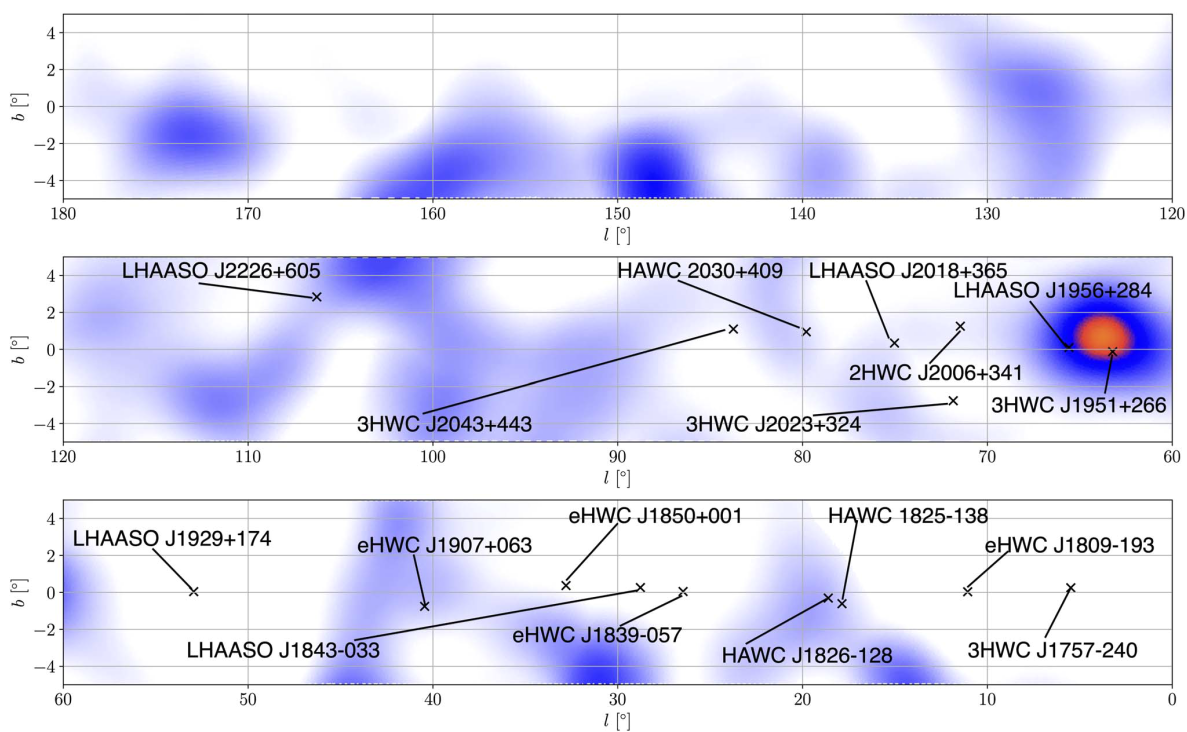}
  \includegraphics[width=\textwidth]{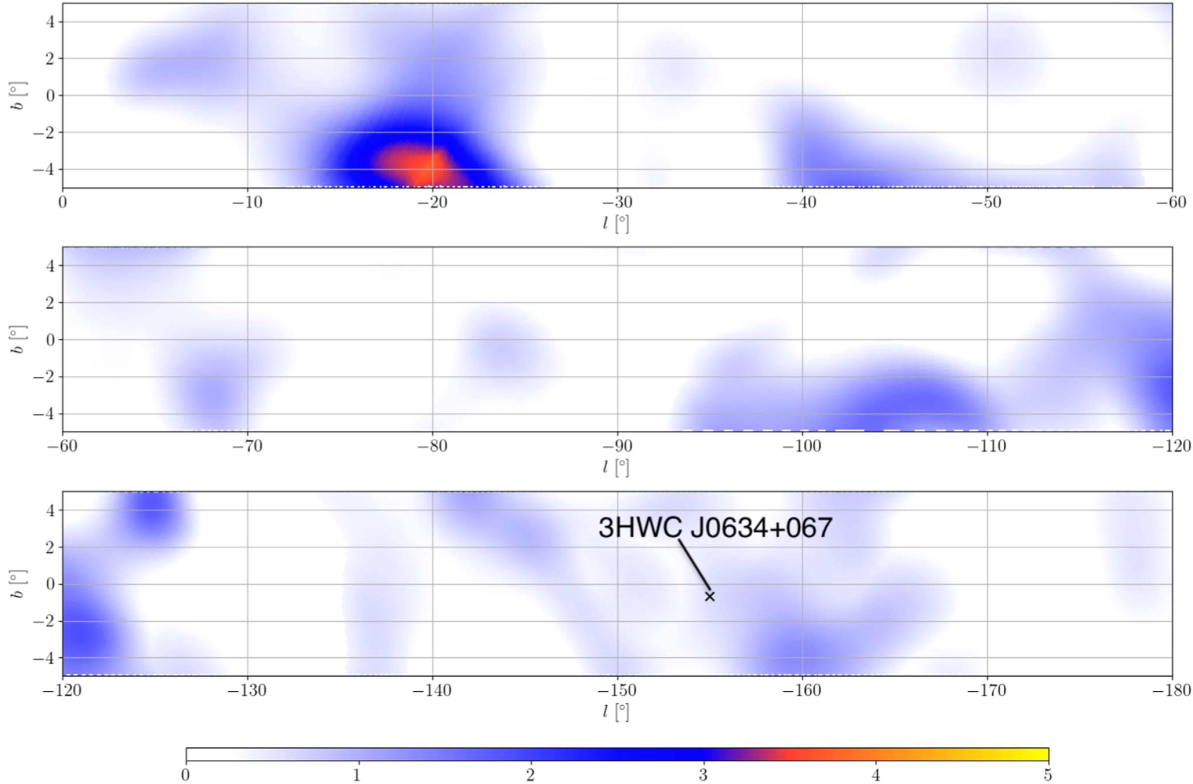}
  \caption{Pre-trial significance map of the scan for extended neutrino sources
  in the galactic plane for a source extension of \SI{2.0}{\degree}.
  Figure taken from \protect\cite{icecube_extended_sources}.}
  \label{fig:extended_significance_map}
\end{figure}

Firstly, a scan is performed for extended sources of predefined angular size in
the galactic plane. The galactic plane is defined as the region with galactic
latitude $\SI{-5}{\degree} \leq b \leq \SI{5}{\degree}$. The source sizes tested
vary between \SI{0.5}{\degree} and \SI{2}{\degree}. No significant
emission was detected in this analysis. The hottest spot is found at the
location of the HAWC source 3HWC J1915+266 (cf. Figure~%
\ref{fig:extended_significance_map}). No significant emission from extended
sources has been detected. The global significance to reject the null hypothesis
is \num{2.6}\,$\sigma$.

Secondly, a catalog search is performed for known \si{TeV} $\gamma$-ray
emitters which have been observed to have an extended morphology. The sources
have been selected from the TeVCat\cite{TeVCat}. They are labelled in Fig.~%
\ref{fig:extended_significance_map}. No source was detected as a significant
neutrino emitter in this analysis. The catalog search resulted in upper
limits on the differential neutrino flux at \SI{50}{TeV}, which could be used
to constrain the hadronic fraction of several sources. %For the strongest
%constraints obtained, see Table~\ref{tab:extended_constraint}.
The most restrictive constraint was obtained for the Cygnus cocoon region, for
which the ratio between the upper limit on the neutrino flux and the expected
neutrino flux, if all observed $\gamma$-ray emission is produced hadronically,
is $\sim \num{0.5}$. For another notable PeVatron candidate, the Boomerang
supernova remnant G106.3+02.7 associated with LHAASO J2226+6057, the upper
limit is a factor of $\sim \num{2.7}$ higher than the neutrino flux expected in
the scenario where all $\gamma$-rays are produced hadronically, indicating the
need for future analyses with improved sensitivity to detect neutrinos from this
source. For a full list of all obtained constraints, see\cite{icecube_extended%
_sources}.

\section{Summary and Outlook}
A diffuse neutrino flux from the galactic plane has been observed with a
significance of \num{4.5}\,$\sigma$ in a model-dependent template search
with a new sample of cascade-like neutrino events. The data currently do not
allow to discriminate between different models of diffuse emission and
discrete sources.

Searches for discrete galactic neutrino sources have so far not resulted in
significant detection. Several searches using samples of track-like neutrino
events have made it possible to put constraints on the hadronic contribution
to the observed $\gamma$-ray emission for several sources.

The results of the different analyses using cascade and track-like events are
consistent with each other. Current analyses work on joint fits of both cascade
and track data.\cite{Fuerst:20236x}

Near-future planned detectors will allow for more detailed searches for
PeVatrons, both kilometer-scale detectors currently constructed in the northern
hemisphere such as KM3Net\cite{km3net_detector} and Baikal-GVD%
\cite{baikal_gvd_detector} as well as the planned extension of
the IceCube neutrino observatory, IceCube-Gen2, which should increase the
effective area by a factor of about 5. Furthermore, multi-messenger analyses
using both neutrino and photon data, together with source-specific
leptohadronic modelling are bound to bring new insights into emission processes
in our galaxy.

%\section*{Acknowledgments}

%This is where one places acknowledgments for funding bodies etc.
%Note that there are no section numbers for the Acknowledgments, Appendix
%or References.

\section*{References}
\bibliography{sandrock}

\begin{thebibliography}{10}

\bibitem{icecube_detector}
M.~G. Aartsen et~al.
\newblock The {IceCube} neutrino observatory: instrumentation and online
  systems.
\newblock {\em J. Instrum.}, 12:P03012, 2017.

\bibitem{icecube_tau}
R.~Abbasi et~al.
\newblock Observation of seven astrophysical tau neutrino candidates with
  {IceCube}.
\newblock {\em Phys. Rev. Lett.}, 132:151001, 2024.

\bibitem{lhaaso_21}
Z.~{Cao} et~al.
\newblock Ultrahigh-energy photons up to 1.4 petaelectronvolts from 12
  $\gamma$-ray galactic sources.
\newblock {\em Nature}, 594(7861):33--36, 2021.

\bibitem{icecube_lhaaso}
R.~Abbasi et~al.
\newblock Searches for neutrinos from {LHAASO UHE} $\gamma$-ray sources.
\newblock {\em Astrophys. J. Lett.}, 945:L8, 2023.

\bibitem{Illuminati:2021Mm}
Giulia Illuminati.
\newblock {Searches for point-like sources of cosmic neutrinos with 13 years of
  ANTARES data}.
\newblock In {\em Proceedings of 37th International Cosmic Ray Conference
  {\textemdash} PoS(ICRC2021)}, volume 395, page 1161, 2021.

\bibitem{icecube_10yr}
M.~G. Aartsen et~al.
\newblock Time-integrated neutrino source searches with 10 years of {IceCube}
  data.
\newblock {\em Phys. Rev. Lett.}, 124:051103, 2020.

\bibitem{ahler_murase2014}
M.~Ahlers and K.~Murase.
\newblock Probing the galactic origin of the {IceCube} excess with gamma rays.
\newblock {\em Phys. Rev. D}, 90:023010, 2014.

\bibitem{icecube_galactic_plane}
R.~Abbasi et~al.
\newblock Observation of high-energy neutrinos from the galactic plane.
\newblock {\em Science}, 380:1338--1343, 2023.

\bibitem{icecube_jinst}
R.~Abbasi et~al.
\newblock A convolutional neural network based cascade reconstruction for the
  {IceCube} neutrino observatory.
\newblock {\em J. Instrum.}, 16:P07041, 2021.

\bibitem{ackermann_2012}
M.~Ackermann et~al.
\newblock {Fermi-LAT} observations of the diffuse $\gamma$-ray emission:
  Implications for cosmic rays and the interstellar medium.
\newblock {\em Astrophys. J.}, 750:3, 2012.

\bibitem{galprop}
A.~W. Strong and I.~V. Moskalenko.
\newblock Propagation of cosmic-ray nucleons in the galaxy.
\newblock {\em Astrophys. J.}, 509:212, 1998.

\bibitem{gaggero_2015}
D.~Gaggero et~al.
\newblock The gamma-ray and neutrino sky: A consistent picture of {Fermi-LAT},
  {Milagro}, and {IceCube} results.
\newblock {\em Astrophys. J. Lett.}, 815:L25, 2015.

\bibitem{gal_optical}
A.~Mellinger.
\newblock A color all-sky panorama image of the milky way.
\newblock {\em Publ. Astron. Soc. Pac.}, 121:1180, 2009.

\bibitem{gal_fermi}
NASA Goddard Space~Flight Center.
\newblock Fermi's 12-year view of the gamma-ray sky.
\newblock https://svs.gsfc.nasa.gov/14090, 2022.

\bibitem{icecube_extended_sources}
R.~Abbasi et~al.
\newblock Search for extended sources of neutrino emission in the galactic
  plane with {IceCube}.
\newblock {\em Astrophys. J.}, 956:20, 2023.

\bibitem{hess_galactic_plane_survey}
H.~Abdalla et~al.
\newblock The {H.E.S.S.} galactic plane survey.
\newblock {\em Astron. \& Astrophys.}, 612:A1, 2018.

\bibitem{hawc_hess_gp_survey}
H.~Abdalla et~al.
\newblock {TeV} emission of galactic plane sources with {HAWC} and {H.E.S.S.}
\newblock {\em Astrophys. J.}, 917:6, 2021.

\bibitem{TeVCat}
S.~P. Wakely and D.~Horan.
\newblock {TeVCat}: An online catalog for very high energy gamma-ray astronomy.
\newblock In {\em Proceedings of the 30th International Cosmic Ray Conference},
  volume~3, page 1341, 2008.

\bibitem{Fuerst:20236x}
P.~Fürst.
\newblock {Galactic and Extragalactic Analysis of the Astrophysical Muon
  Neutrino Flux with 12.3 years of IceCube Track Data}.
\newblock In {\em Proceedings of 38th International Cosmic Ray Conference
  {\textemdash} PoS(ICRC2023)}, volume 444, page 1046, 2023.

\bibitem{km3net_detector}
S~Adri{\'a}n-Mart{\'\i}nez et~al.
\newblock Letter of intent for {KM3NeT} 2.0.
\newblock {\em J. Phys. G Nucl. Part. Phys.}, 43(8):084001, August 2016.

\bibitem{baikal_gvd_detector}
A~V Avrorin et~al.
\newblock Deep-underwater {Cherenkov} detector in lake {Baikal}.
\newblock {\em J. Exp. Theor. Phys.}, 134(4):399--416, April 2022.

\end{thebibliography}
\end{document}